\documentclass[techreport,review,12pt]{elsarticle}
\makeatletter
\def\ps@pprintTitle{%
  \let\@oddhead\@empty
  \let\@evenhead\@empty
  \let\@oddfoot\@empty
  \let\@evenfoot\@oddfoot
}
\makeatother

\usepackage{graphicx}
\usepackage{subfigure} 

\usepackage{amsmath}
\usepackage{amssymb}

\usepackage{amsthm,amsfonts}
\usepackage{bm}

\usepackage{lineno}

\usepackage{natbib}

\usepackage{algorithm}
\usepackage{algorithmic}

\usepackage{hyperref}
\usepackage{diagbox}

\usepackage{tikz}
\usetikzlibrary{arrows,fit,positioning}
\tikzstyle{block}=[draw opacity=0.7,line width=1.4cm]

\begin{document}
\setlength{\belowdisplayskip}{7.5pt} \setlength{\belowdisplayshortskip}{0pt}
\setlength{\abovedisplayskip}{7.5pt} \setlength{\abovedisplayshortskip}{0pt}

\begin{frontmatter}
%\linenumbers



\title{Game Theoretic Consequences of Resident Matching}


%\author{Yue Wu}\ead{yueswu@uw.edu}
\author{Yue Wu\corref{cor1}}
\cortext[cor1]{Corresponding author}
\ead{yueswu@uw.edu}
\address{University of Washington}

% limit 100 words abstract - currently 112
\begin{abstract}
The resident matching algorithm, Gale-Shapley, currently used by SF Match and the National Resident Match Program (NRMP), has been in use for over 50 years without fundamental alteration. 
The algorithm is a `stable-marriage' method that favors applicant outcomes. 
However, in these 50 years, there has been a big shift in the supply and demand of applicants and programs. 
These changes along with the way the Match is implemented have induced a costly race among applicants to apply and interview at as many programs as possible. 
Meanwhile programs also incur high costs as they maximize their probability of matching by interviewing as many candidates as possible. 
\end{abstract}

%%FIXME
\end{frontmatter}

%\linenumbers
\allowdisplaybreaks

\section{Introduction} \label{sec:intro}
The tech report is organized as follows: 
Section~\ref{sec:background} Background,
Section~\ref{sec:matchingAlgos} Matching Algorithm history and details,
Section~\ref{sec:ResMatching} Resident matching with no costs, 
Section~\ref{sec:ResMatchingWithCosts} Resident matching with costs.

\section{Background} \label{sec:background}
Residency matching connects graduating medical students with residency programs as they pursue further medical training and specialization.
Each year, over 43,000 US and international graduates apply for 31,000 residency positions. 
Unique to medicine, an algorithm determines the fate of graduating students. 

The residency matching algorithm has evolved over time. 
The first residency programs, then called internships, were introduced in the 1920s as optional postgraduate training. 
These programs initially attracted only a few medical graduates, who wished to gain more exposure to clinical medicine. 
The inadequate supply of interns led to fierce competition among the programs, which manifested as a race between programs to secure binding commitments from potential graduates as early as possible \cite{Roth1984-rx}. 
This resulted in medical students receiving internship offers up to two years before graduation \cite{Turner1945-dp}.
This process not only disrupted the last two years of medical school, but was suboptimal on both sides. 
Students often had to accept or reject offers without having heard back from all programs, to which they applied, including more preferred ones. 
Programs had incomplete information and little medical school performance to judge applicant qualifications when extending offers this early.

To counter this, Turner proposed and the medical schools agreed in 1945 to not release transcripts nor give letters of recommendation until a later date. 
However, this did not fundamentally change the process, but simply pushed back the process until the last year of medical school. 
The programs and applicants still played the same adversarial game, but on a shorter time scale. 
The programs made time-limited offers, while the students delayed accepting until hearing back from more preferred programs. 
This led programs to steadily decrease the offer time limits to reduce the risk of their preferred candidates successfully waiting and accepting alternate offers. 
Therefore the programs decreased it to less than 12 hours in 1950, from an initially agreed offer time limit of 10 days in 1945 \cite{Springs1950-aj}. 

In order to avoid this race to the bottom between programs, the National Interassociation Committee on Internships (NICI) was formed in 1950 to examine existing matching plans and perform a trial run for a centralized match system. 
Then in October 1951, 79 medical schools formed the National Student Internship Committee (NSIC) and decided to adopt the Boston Pool Plan \cite{Mullin1951-ei} nationally upon the recommendations of the National Interassociation Committee on Internships (NICI). 
This method was modified to be more equitable to applicants, and then first used nationally in April 1952 \cite{Mullin1952-ch}. 
The National Internship Matching Program (NIMP), now the National Resident Matching Program (NRMP), was incorporated on Jan 7, 1953, to manage and administer the matching process. 

The NRMP algorithm saw only minor changes from 1952 until 1997 \cite{Gusfield1989-qk}. 
It was a stable-marriage algorithm with the programs being the proposing side. 
Then in 1997, after a year of preliminary study \cite{noauthor_undated-vb}, the NRMP decided to adopt an applicant-proposing version of the old algorithm \cite{Roth1999-lc}. 
In ophthalmology, the match process is administered by SF Match, which has used an applicant proposing stable-marriage algorithm since 1977.

In the following section, we discuss the matching algorithms in more detail to better understand their strengths and weaknesses.

\section{Matching Algorithms} \label{sec:matchingAlgos}
\subsection{Boston Pool Plan} \label{sec:bostonPoolPlan}

The initial NIMP algorithm matched the applicants to programs according to the confidential ranked lists submitted by all participants \cite{Mullin1951-ei}. 
The applicants rank all the programs they would be willing to attend in ascending ordinal order, with 1 being most preferred.
The programs group the applicants by tiers and then rank within tiers. 
For example, if a program had n available positions, its first choice applicants would be tier 1 and ranked within tier from one to n, and its next tier students would be ranked n+1 to 2n, and so on. 
The algorithm starts by matching students and programs that each other 1 or most preferred. 
Then, the programs, which had not filled their positions, would be matched with students in their second tier, but who had ranked the programs 1.
If still unfilled, the programs would be next matched with applicants in their second tier and who had ranked the programs 2.
The process continues iteratively and the can be represented as paired ranks in the following order [(1,1), (2,1), (1,2), (2,2), (3,1), (3, 2), (1, 3), (2, 3), (3, 3), …], where the ordered rank pair is the program tier and student rank respectively.

Immediately after this initial NIMP algorithm was proposed, objections arose as students showed that it might penalize highly qualified students. 
For example, suppose an applicant did not match his first choice, and his second choice program had ranked him in the first tier. 
However, given the order of the matching, his second choice program might already have filled all its positions with candidates in its second tier. 
This leads to a situation where both the applicant and the program prefer each other, but are not matched, and is an example of an unstable match.

\subsection{Stable-Marriage Algorithm} \label{sec:galeShapley}

The final adopted NIMP algorithm was modified to avoid these types of situations by using deferred matching \cite{Mullin1952-ch}. 
It is a stable-marriage algorithm, and is mathematically equivalent to the Gale-Shapley algorithm, which won its authors the Nobel Prize in Economics in 2012. 

The canonical stable-marriage algorithm finds stable matches between a group of n men and a group of n women, that have ranked each other in terms of preference. 
A match is stable when for a match tuple of man and woman, denoted (A, B):
\begin{enumerate}
\item there is no woman D that man A prefers, and who also prefers A to her current match
\item there is no man C that woman B prefers, and who also prefers B to his current match
\end{enumerate}
Alternatively a match (A, B) is not stable if:
\begin{enumerate}
\item there is a woman D that man A prefers, and who also prefers A to her current match
\item Or there is man C that woman B prefers, and who also prefers B to his current match
\end{enumerate}

The stable-marriage problem applies not only to matching men and women, but to many other two-sided stable matching problems. 
The resident matching problem can be reworded as a stable marriage problem with the applicants as one side and the residency programs the other. 
The only difference is that residency programs can match as many applicants as they have residency positions. 

For an equal number of participants on each side, who have ranked every potential partner, it was proved by Gale and Shapley that a stable match women exist \cite{noauthor_2011-vv}, 
and their eponymous algorithm finds a stable solution. 
Gale and Shapley had originally applied their algorithm to matching colleges with students, but this is mathematically equivalent to matching programs with medical students.

The Gale-Shapley algorithm takes as input the rank lists of all the participants along with a proposing side. 
Without loss of generality, assume for the following discussion that programs are the proposing side. 
Then the algorithm first selects a program at random from the pool of programs with unfilled positions. 
The program will first propose to its most preferred candidate according to its rank list. 
If that candidate is unmatched, then a tentative match is formed with the program, and the algorithm picks another program to start proposing. 
If the candidate is matched, the algorithm checks if they would prefer the proposing program over their currently matched program. 
If they prefer the proposing program, then their previous match is annulled and they are matched tentatively with the proposer, and their previous partner is added back to the proposing pool. 
The algorithm continues until all program positions have been filled.

\section{Resident matching with no costs} \label{sec:ResMatching}
Consider the general resident matching setup with $N$ applicants, $P$ programs with each program having ${s_{1}, ..., s_{P}}$ available spots, such that there is a total of $S= \sum_{i=i}^{p}s_{i}$ available spots. 
In the current process, the applicants apply to the programs, and are then either invited for an interview or rejected.
The applicants then rank the programs in strict ordinal order in terms of their preference.
On the other side, the programs also rank the applicants, specifically their invited interviewees in ordinal preference as well.
Finally, applicants' and programs' rank lists are matched using a stable-marriage algorithm favouring the applicant.

First, we show that in the current resident matching process the best strategy for applicants is to rank as many as programs as possible when there is no cost to playing the ranking game.
We prove this by induction starting with a sub-game with one applicant, $A$, and one program $\alpha$, at the ranking stage of the resident matching process.
Then the actions of the applicant for this sub-game are either rank or not rank the program and vice versa for the program.
The payout scenario is shown below: 
\begin{center}
\begin{tabular}{ c | c | c }
\backslashbox{Applicant $A$}{Program $\alpha$} 		& rank 														& not rank \\ 
\hline
 rank 										& $f_{A}(r_{A \alpha}, M)$, $g_{\alpha}(\rho_{\alpha A}, M)$ 		& $f_{A}(r_{A \alpha}, M)$, 0 \\  
\hline
 not rank 									& 0, $g_{\alpha}(\rho_{\alpha A}, M)$									& 0, 0
\end{tabular}
\label{tab:2player_subgame}
\end{center}
This two-by-two payout table has applicant actions as rows and program actions columns.
The payout is a 2-tuple of applicant payout and program payout in that order.

Clearly, if neither ranked the other then the payoff for both is zero.
Not ranking in this sub-game is equivalent to non-participation.
Let $r_{A \alpha}$ denote the rank applicant $A$ assigned program $\alpha$.
Similarly, let $\rho_{\alpha A}$ be the rank program $\alpha$ gave applicant $A$.
Next let $f_{A}$ and $g_{\alpha}$ denote the payout functions of applicant $A$ and program $\alpha$ respectively. 
For clarity, the payout $f_{A}$ of the candidate is a function of the candidate's rank of the program, denoted $r_{A \alpha}$, and the match status $M=(0,1)$.
Conversely, the payout $g_{\alpha}$ of the program is function of the program's rank of the applicant, denoted $\rho_{\alpha A}$, and the match status $M=(0,1)$.
%
%We introduce the payout function notation now, as it generalizes to games with more participants.
The payout functions should increase monotonically with decreasing rank of match, as the participants' payouts should increase when they match with more preferred partners:
%In addition, the payouts should increase monotonically so that no payout for a less preferred partner exceeds that of a more preferred one:
\begin{align}
	r 	& \in \mathbb{Z^{+}}	\\
	M 	&\in (0, 1)	\\
	f: 	&(\mathbb{Z^{+}}, (0,1))  \to \mathbb{R} \\
	g: 	&(\mathbb{Z^{+}}, (0,1))  \to \mathbb{R} \\
	\text{If } &r_{2} 		> r_{1}, \\ 
	\text{then } &f(r_{1}, M)  \geq f(r_{2}, M), \\
	\text{and } &g(r_{1}, M)  \geq g(r_{2}, M) 
\end{align}
Next note the match status $M=(0,1)$, represents whether the participant matched (1) or not (0).
Thus, the payouts can also be written as:
\begin{align}
	f_{A}(r_{A \alpha}, M) &= f_{A}(r_{A \alpha}, M=0) + f_{A}(r_{A \alpha}, M=1)	\\
	g_{\alpha}(\rho_{\alpha A}, M) &= g_{\alpha}(\rho_{\alpha A}, M=0) + g_{\alpha}(\rho_{\alpha A}, M=1)
\end{align}
Clearly, if the participant matched, $M=1$, then the payout is positive, as they found a partner.
If they did not match, $M=0$, their payout is at worst zero, as their status has not changed and they have not lost anything by participating in the current game, where there are no participation costs.
However, it is possible that the payouts are non-negative even if a match did not occur. 
For example, in the one student and one program sub-game, the applicant or program might find it rewarding to know if the counter-party ranked them or not.
In practice, the payout for this information discovery should be intuitively much lower than for matching.
%TODO -We can formally articulate this by putting a distribution on the payout of not matching relative to relative to the payout of matching.
In this sub-game, if both the applicant and program ranked each other, they will be matched under the stable-marriage algorithm; $M=1$ for both.
However, if only side ranked the other there will be no match under the stable-marriage algorithm; $M=0$.
Therefore the payout table simplifies to:
\begin{center}
\begin{tabular}{ c | c | c }
\backslashbox{applicant $A$}{program $\alpha$} 		& rank 														& not rank \\ 
\hline
 rank 										& $f_{A}(r_{A \alpha}, M=1)$, $g_{\alpha}(\rho_{\alpha A}, M=1)$ 		& $f_{A}(r_{A \alpha}, M=0)$, 0 \\  
\hline
 not rank 									& 0, $g_{\alpha}(\rho_{\alpha A}, M=0)$									& 0, 0
\end{tabular}
\label{tab:2player_subgame_simplified}
\end{center}
Thus, for the one applicant and one program sub-game, the highest payout is achieved when both rank each other, as $f_{A}(r_{A \alpha}, M=1) \geq f_{A}(r_{A \alpha}, M=0) \geq 0$ and $g_{\alpha}(\rho_{\alpha A}, M=1) \geq g_{\alpha}(\rho_{\alpha A}, M=0) \geq 0$ 	.

Next, suppose the sub-game was extended to have two programs, $\alpha$ and $\beta$, but still one participant $A$. 
Then there will be two payout tables.
First, there is the payout table for applicant $A$ and program $\alpha$:
\begin{center}
\begin{tabular}{ c | c | c }
\backslashbox{Applicant $A$}{Program $\alpha$} 		& rank 														& not rank \\ 
\hline
 rank 										& $f_{A}(r_{A \alpha}, M)$, $g_{\alpha}(\rho_{\alpha A}, M)$ 		& $f_{A}(r_{A \alpha}, M)$, 0 \\  
\hline
 not rank 									& 0, $g_{\alpha}(\rho_{\alpha A}, M)$									& 0, 0
\end{tabular}
\label{tab:1app_2prog_subgame_aa}
\end{center}
Then there is the payout for $A$ versus program $\beta$
\begin{center}
\begin{tabular}{ c | c | c }
\backslashbox{Applicant $A$}{Program $\beta$} 		& rank 														& not rank \\ 
\hline
 rank 										& $f_{A}(r_{A \beta}, M)$, $g_{\beta}(\rho_{\beta A}, M)$ 		& $f_{A}(r_{A \beta}, M)$, 0 \\  
\hline
 not rank 									& 0, $g_{\beta}(\rho_{\beta A}, M)$									& 0, 0
\end{tabular}
\label{tab:1app_2prog_subgame_ab}
\end{center}
The difference in payouts between this game and the previous game ~\ref{tab:2player_subgame_simplified} is that there is no guarantee the applicant and a program match even if they rank each other.
The total applicant payout is the sum of applicant payouts for each program:
\begin{align}
	f_{A} &= \max{[f_{A}(r_{A \alpha}, M), 0]} + \max{[f_{A}(r_{A \beta}, M), 0]} \geq 0 \label{eq:app_payout_progs}
\end{align}
The payout in \eqref{eq:app_payout_progs} is the maximum of a non-negative payout function and zero depending on whether the applicant ranked a program or not.
Clearly, the applicant should rank both programs no matter what the programs do.
Furthermore, when the applicant ranks both programs, Equation~\eqref{eq:app_payout_progs} can be expanded as:
\begin{align}
	f_{A} &= f_{A}(r_{A \alpha}, M=0) + f_{A}(r_{A \alpha}, M=1) + f_{A}(r_{A \beta}, M=0) + f_{A}(r_{A \beta}, M=1) \label{eq:app_payout_progs_expanded}
\end{align}
Then if neither program ranked the applicant, this simplifies to:
\begin{align}
	f_{A} &= f_{A}(r_{A \alpha}, M=0) + f_{A}(r_{A \beta}, M=0) \geq 0	\label{eq:app_payout_progs_neither}
\end{align}
Next, if either program $\alpha$ or $\beta$ but not both ranked applicant $A$, then $A$ will match with the program that ranked $A$.
Without loss of generality suppose only $A$'s less preferred program $\beta$ ranked $A$, then the payout is:
\begin{align}
	f_{A} &= f_{A}(r_{A \alpha}, M=0) + f_{A}(r_{A \beta}, M=1) \geq \eqref{eq:app_payout_progs_neither} \geq 0 \label{eq:app_payout_progs_beta} \\
	\text{Since, } & f_{A}(r_{A \beta}, M=1) > f_{A}(r_{A \beta}, M=0) \nonumber
\end{align}
Finally, if both programs ranked $A$, $A$ would match the more preferred program $\alpha$ and the payout is:
\begin{align}
	f_{A} &= f_{A}(r_{A \alpha}, M=1) + f_{A}(r_{A \beta}, M=0) \geq \eqref{eq:app_payout_progs_beta} \geq 0 \label{eq:app_payout_progs_alpha} \\
	\text{Since, } & f_{A}(r_{A \alpha}, M=1) \geq f_{A}(r_{A \beta}, M=1) \nonumber \\
	\text{and, } & f_{A}(r_{A \alpha}, M=0) \geq f_{A}(r_{A \beta}, M=0) \nonumber
\end{align}
Clearly, the applicant achieves the highest payout by ranking both programs while holding the actions of the programs constant.

For the programs the payouts are $\max{[g_{\alpha}(\rho_{\alpha A}, M), 0]}$ and $\max{[g_{\beta}(\rho_{\beta A}, M), 0]}$ respectively. 
Then the optimal for the programs is to rank the applicant, because in the worst case they do not match but $g_{\alpha}(\rho_{\alpha A}, M=0) \geq 0$ and $g_{\beta}(\rho_{\beta A}, M=0) \geq 0$.

Next, consider the more common case in resident matching where applicants outnumber program spots.
For this scenario, consider the subgame with two applicants $A$ and $B$ and one program $\alpha$.
This is similar to the game with one applicant and two programs.
The difference is that the payouts of the applicants are:
\begin{align}
	f_{A} &= \max{[f_{A}(r_{A \alpha}, M), 0]} \geq 0  \nonumber \\  
	f_{B} &= \max{[f_{B}(r_{B \alpha}, M), 0]} \geq 0  \label{eq:app_payout_apps}
\end{align}
Again both applicants will achieve higher payout by ranking versus not ranking the program, no matter what the program or the other applicant does.
Similarly, program $\alpha$'s payout is:
\begin{align}
	f_{A} &= \max{[g_{\alpha}(\rho_{\alpha A}, M), 0]} + \max{[g_{\alpha}(\rho_{\alpha B}, M), 0]} \geq 0	\label{eq:prog_payout_apps}
\end{align}
Clearly, the program achieves higher payout by ranking both applicants.
Then the program either does not match at all when neither applicant ranked program or it matches one of the applicants.
All these scenarios have payouts no worse than not ranking either applicant.

We can extend the subgames to include more applicants and programs.
In general, when there are $N$ applicants and $P$ programs, the payout of applicant $i$'s is:
\begin{align}
	f_{i} &= \sum_{p=1}^{P} \max{[f_{i}(r_{i p}, M), 0]} \geq 0 \label{eq:app_payout_general}
\end{align}
Every term of the summation in Equation \eqref{eq:app_payout_general} corresponds to the payoff for ranking a program.
Clearly, since every term is non-negative it is better for applicant $i$ to always rank a program.
Therefore for applicant $i$, there is a pure strategy where he or she ranks all the programs that maximizes his or her payout.
Similarly, since the payouts for the other applicants $j \neq i$, are similar to Equation \eqref{eq:app_payout_general}, the other applicants will also maximize their payout by ranking all programs.

%So no matter what the other participants do, applicant $i$ will do no worse when he or she ranks all the programs.
%Intuitively, this make sense as in this game with no costs, applicants will do no worse by ranking the $(p+1)^{th}$ program compared to just ranking $p$ programs, as $\max{[f_{i}(r_{i, p+1}, M), 0]} \geq 0$.
In practice, most applicants and programs will not match against each other so that the pairwise payoff for applicant $i$ and program $p$ is likely $f_{i}(r_{i p}, M=0),  g_{i}(\rho_{p i}, M=0)$.
Next note that under the stable-marriage Gale-Shapley algorithm \cite{noauthor_2011-vv}
%\ref{fig:gale_shapley} \cite{gale_shapley}
, applicants will match with their highest ranked program that also preferred them to other candidates, until the number of spots at the programs are filled.
So for applicants that matched there will be a payoff tuple with $M=1$, e.g. $f_{i}(r_{i p}, M=1),  g_{i}(\rho_{p i}, M=1)$.
%\begin{figure}
%	\centering 
%	\includegraphics[scale=.65]{gale_shapley.png} 
%	\caption{Gale-Shapley Algorithm}
%	\label{fig:gale_shapley}
%\end{figure}

%Next, is it possible that an applicant to improve on that match by manipulating his or her rank list, so that they match with a more preferred program, knowing all other rank lists?
%\cite{Roth} showed that lying is not optimal under stable-marriage Gale-Shapley algorithm. 
%\textbf{Understand this proof better, because imagine case where applicant peeks all other rank lists and conclude that they would at least match if they put their worst ranked choice at the top, then this is better for them. However, Roth shows the probability of this happening is vanishingly small.}

Now consider the program payouts.
The payout for a program $\iota$ is:
\begin{align}
	g_{\iota} &= \sum_{i=1}^{N} \max{[g_{\iota}(\rho_{\iota i}, M), 0]} \geq 0 \label{eq:sch_payout_general}
\end{align}
Just like the applicant case, it is optimal for program $\iota$ to rank all applicants.
By the same logic, it is optimal for all programs to rank all applicants.

Therefore, in a framework where applicants and programs can take the actions a) rank or b) not rank without incurring any costs, it is optimal for the applicants and programs to rank every potential partner.
They should rank a potential partner even if there is no guarantee that that partner ranked them.
This framework with no costs is unrealistic, but introduces the tools in Section~\ref{sec:ResMatchingWithCosts}.
The no cost frame is unrealistic because:
\begin{enumerate}
	\item There is full knowledge all potential partners. This may be true for medical students knowing all potential residency programs, but the converse is not true. 
	Clearly, residency programs only know the students that applied to their program.
	\item Not all applicants apply to all programs because of i) costs of applications or ii) payoff of matching with certain programs is 0.
	\item There is an interview process where applicants and programs identify more likely partners. This interview process serves as mutual signalling, but has high costs.
\end{enumerate}
%In the next section \ref{sec:ResMatchingWithCosts}, we consider the optimality of ranking and not ranking with costs.

\section{Resident matching with costs}\label{sec:ResMatchingWithCosts}
As shown in the previous section \ref{sec:ResMatching}, it is always better to rank.
There are no costs to ranking for an applicant.
They can rank all potential residencies online.
However, the applicant ranks only have reasonable potential payoff if the counterparty likely ranked them.
For example, if an applicant ranked a program that they did not even apply to, there is almost no chance that the  program ranked the applicant as they would not be aware of the applicant's existence.
Similarly, an applicant can rank a program that they applied to but were rejected for interview.
Again it is unlikely, the program ranked a rejected applicant or at least not likely that they ranked a rejected applicant highly enough that they potentially matched.
Therefore, to increase the odds of a better potential match and thereby potential payoff,  an applicant should increase the odds of being ranked by the programs. 
They can start by applying to as many programs as possible, resources permitting.
The application cost is a step-wise function of the number of programs applied, \href{https://residency.wustl.edu/application/cost-of-applying/}{Ophthalmology application costs}.
These costs range from \$60 for the first ten programs to \$35 per program after 40 programs.
%If a student applied to all 115 programs in Opthalmology, the total application fee will amount to \$3135.
%However, these costs are small relative the potential payout of matching against a program and becoming a trained ophthalmologist.
%The AMA annual average income for an ophthalmologist was \$325,000.
After an applicant passes the first round filter and is invited for an in-person interview, there are the costs of attending interviews.
This cost will vary depending on the applicant home city and the program city.
Finally, every applicant will have different monetary and time budgets for the whole application and interview process.

Incorporating these costs and budgets into a generic applicant $i$'s payout in Equation \eqref{eq:app_payout_general}:
\begin{align}
	f_{i} = & \sum_{p=1}^{q_{i}} \left( \mathrm{E}_{p} ( f_{i}(r_{i p}, M)] - A(p) \right) \label{eq:app_payout_general_with_cost} \\
	\mathrm{E}_{p} [f_{i}(r_{i p}, M)] = & ( p(I_{p}=1) \times \mathrm{E}[f_{i}(r_{i p}, M) | I_{p}=1] - \mathrm{E}[C_{p}] ) +  \nonumber \\ & p(I_{p}=0) \times \mathrm{E} [f_{i}(r_{i p}, M)|I_{p}=0] \label{eq:interview_payout} \\ 
	\text{subject to } & \sum_{p=1}^{q_{i}} (p(I_{p}=1)\mathrm{E}[C_{p}] + A(p) ) \leq B_{i} \, \nonumber \\ 
	\text{and } &\sum_{p=1}^{q_{i}} t_{p} \leq T_{i}  \nonumber
\end{align}
Equation~\eqref{eq:app_payout_general_with_cost} states that the payout of the whole process for applicant $i$ is the sum of the expected payoff for applying to $q_{i}$ programs. 
The payout for applying to the $p^{th}$ program is the expected payout 	$\mathrm{E}_{p} [f_{i}(r_{i p}, M)]$ minus the cost of application $A(p)$.
The term $A(p)$ is the cost for applying to the $p^{th}$ program and is a deterministic step function in resident matching. 
The expectated payoff of applying to the $p^{th}$ program \eqref{eq:interview_payout} is the payoff for being accepted for an interview, $p(I_{p}=1)$, multiplied the expected payoff for attending the interview, $\mathrm{E} [f_{i}(r_{i p}, M)|I_{p}=1]$, minus the expected cost of attending the interview $\mathrm{E}[C_{p}]$ plus the payout if not accepted for interview. 
The number of applications and interviews applicant $i$ can pay for is constrained by the monetary budget $B_{i}$ and the time budget $T_{i}$.

\begin{figure}
	\centering 
	\includegraphics[scale=.25]{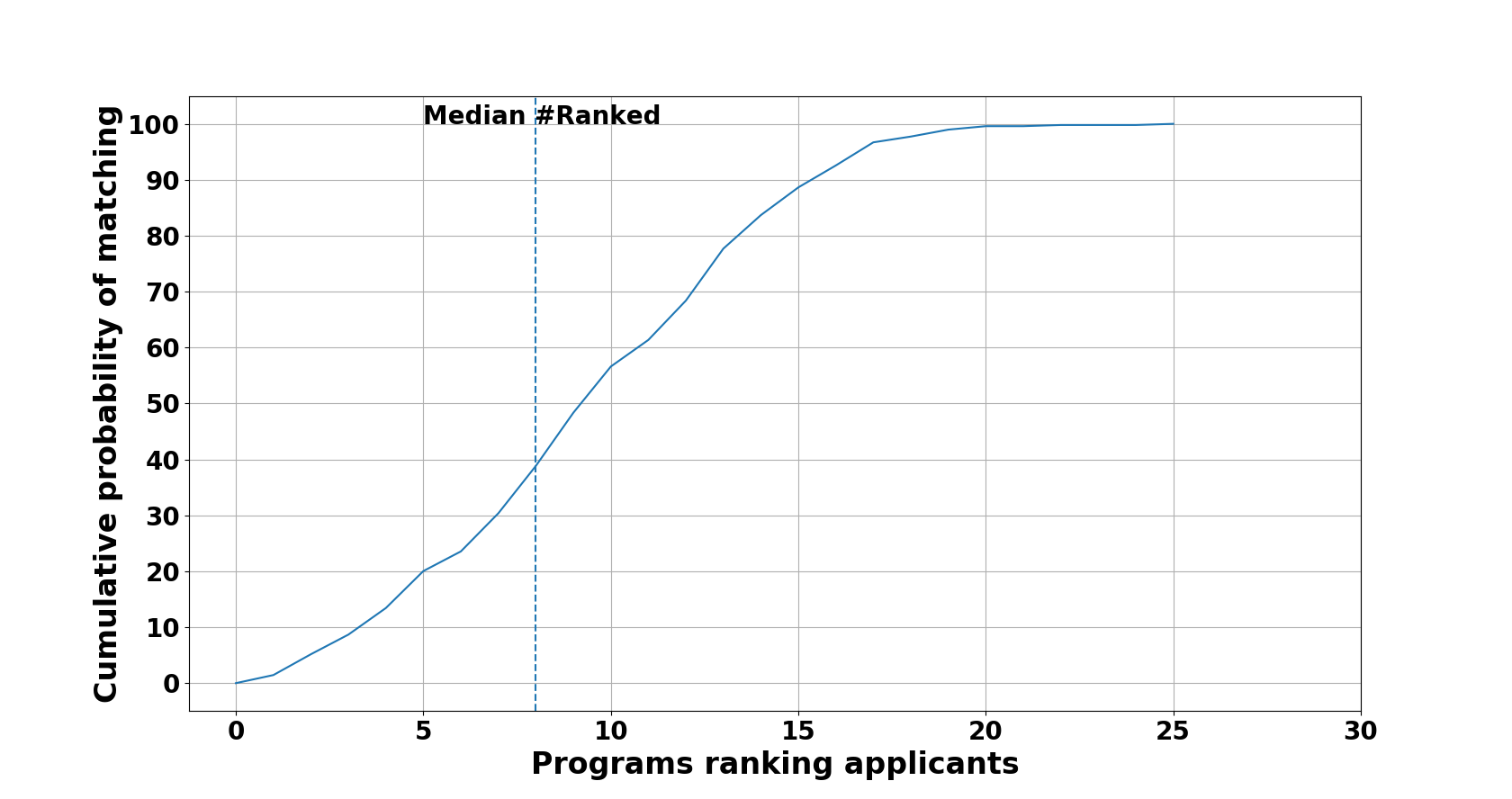} 
	\caption{Cumulative probability of matching for number of times applicant was ranked}
	\label{fig:matching_cum_prob_numRanked_app_by_sch}
\end{figure}

\begin{figure}[H]
	\centering 
	\includegraphics[scale=.25]{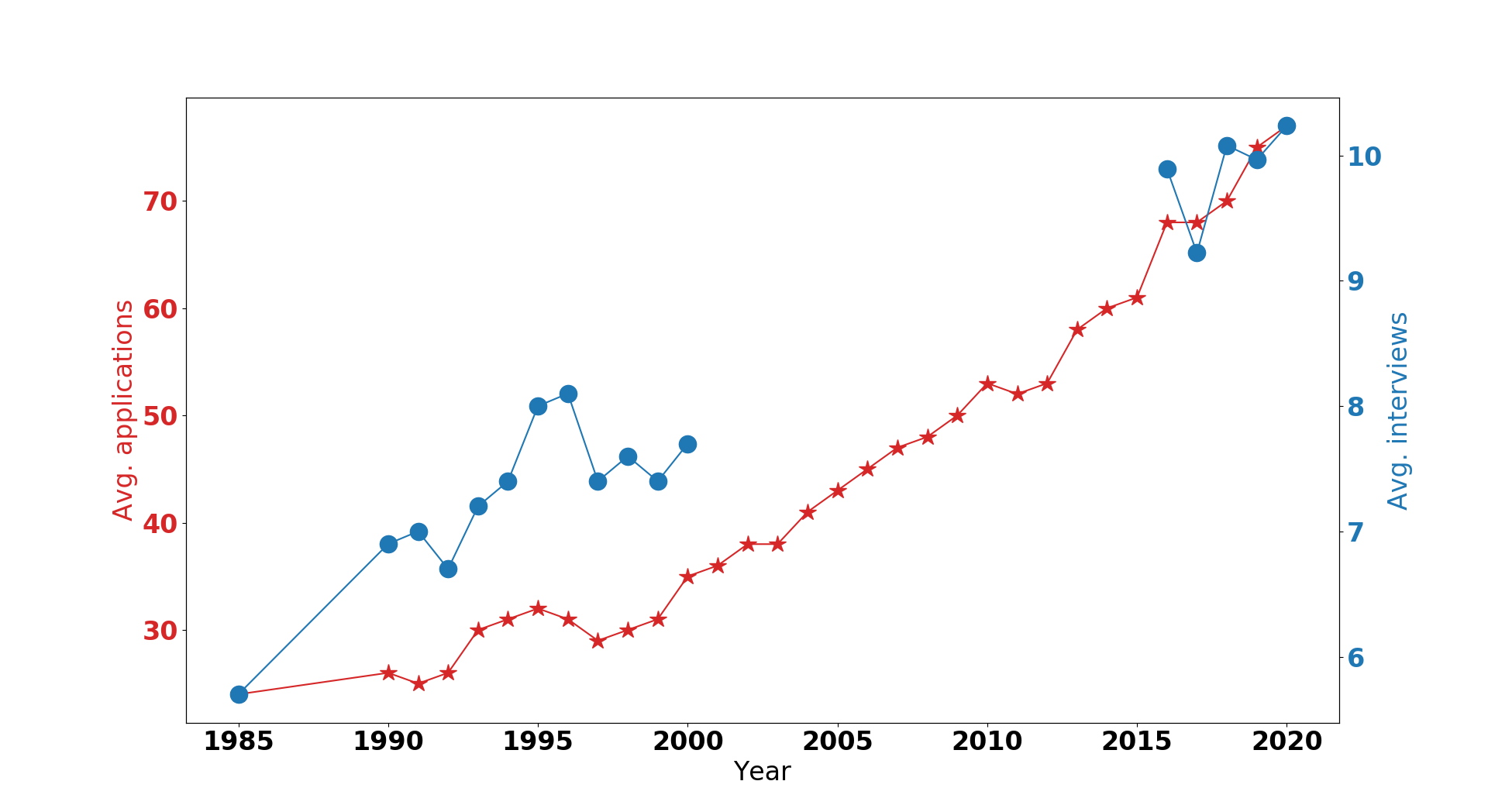} 
	\caption{Average number of Ophthalmology applications and interviews over time}
	\label{fig:trend_app_interviews_v2}
\end{figure}

\begin{figure}[H]
	\centering 
	\includegraphics[scale=.5]{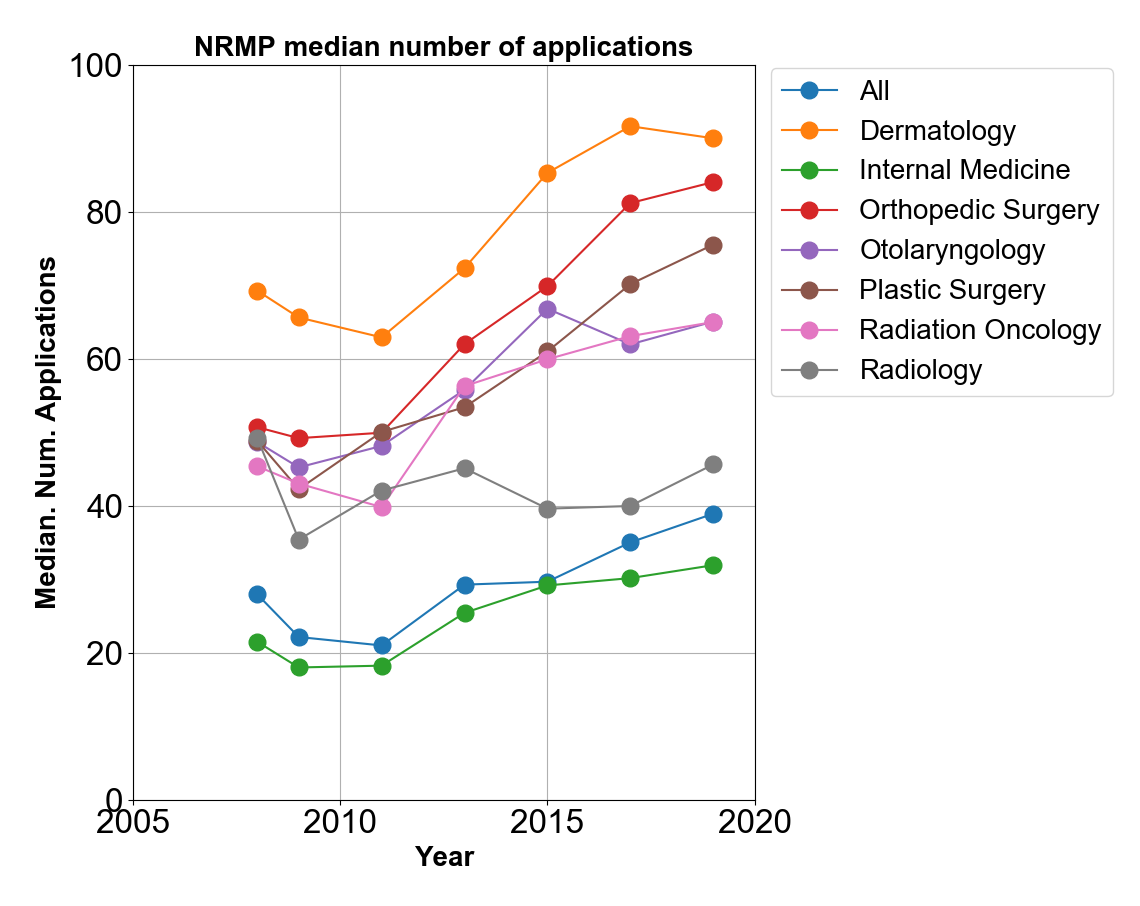} 
	\caption{Median number of NRMP applications over time}
	\label{fig:nrmp_median_applications_by_year}
\end{figure}

First, we analyse the effect of applying to programs.
Each term in Equation~\eqref{eq:interview_payout} consists of the probability of being accepted for an interview, 
and the expected payoff being ranked by program if interviewed or not interviewed.
We use Bayesian payoffs to estimate the payoff given empirical data \cite{Moore2018-pv}.  
The probability of matching is incremental with each extra interview increasing the probability of matching to a program.
The empirical cumulative distribution of matching for an applicant in Ophthalmology given the number of interviews is given in 
Figure~\ref{fig:matching_cum_prob_numRanked_app_by_sch}.
The probability of being accepted for an interview is approximately $1/7 \approx 14\% $ in SF Match data.
Now the cost for applying to all 116 Ophthalmology programs in 2019 is \$3170.
Applying to all 116 programs is expected to yield 16.6 interviews.
The cost per interview was estimated to be \$404.
The total cost of applying to all and then attending the 16.6 expected interviews is \$9,865.
This pales in comparison to the estimated annual average income of \$366,000 for an ophthalmologist in 2019 \cite{noauthor_undated-zz},
and an expected career income of over \$10,000,000, 
even discounted for present value.
The only thing that should stop applicants applying to all programs is current budget constraints,
but \$9,865 should not be the limiting factor for most applicants given the overall cost of medical education.
This is borne out by the trend in Ophthalmology applications in Figure~\ref{fig:trend_app_interviews_v2}.
This trend is not limited to Ophthalmology, and is also seen in the NRMP specialties in Figure~\ref{fig:nrmp_median_applications_by_year}.
In some of these specialties, applicants are applying to 90\% of all programs as seen in Figure 1c. of \citet{wu2021inefficiencies}, reproduced below:
\begin{figure}[H]
	\centering 
	\includegraphics[scale=.5]{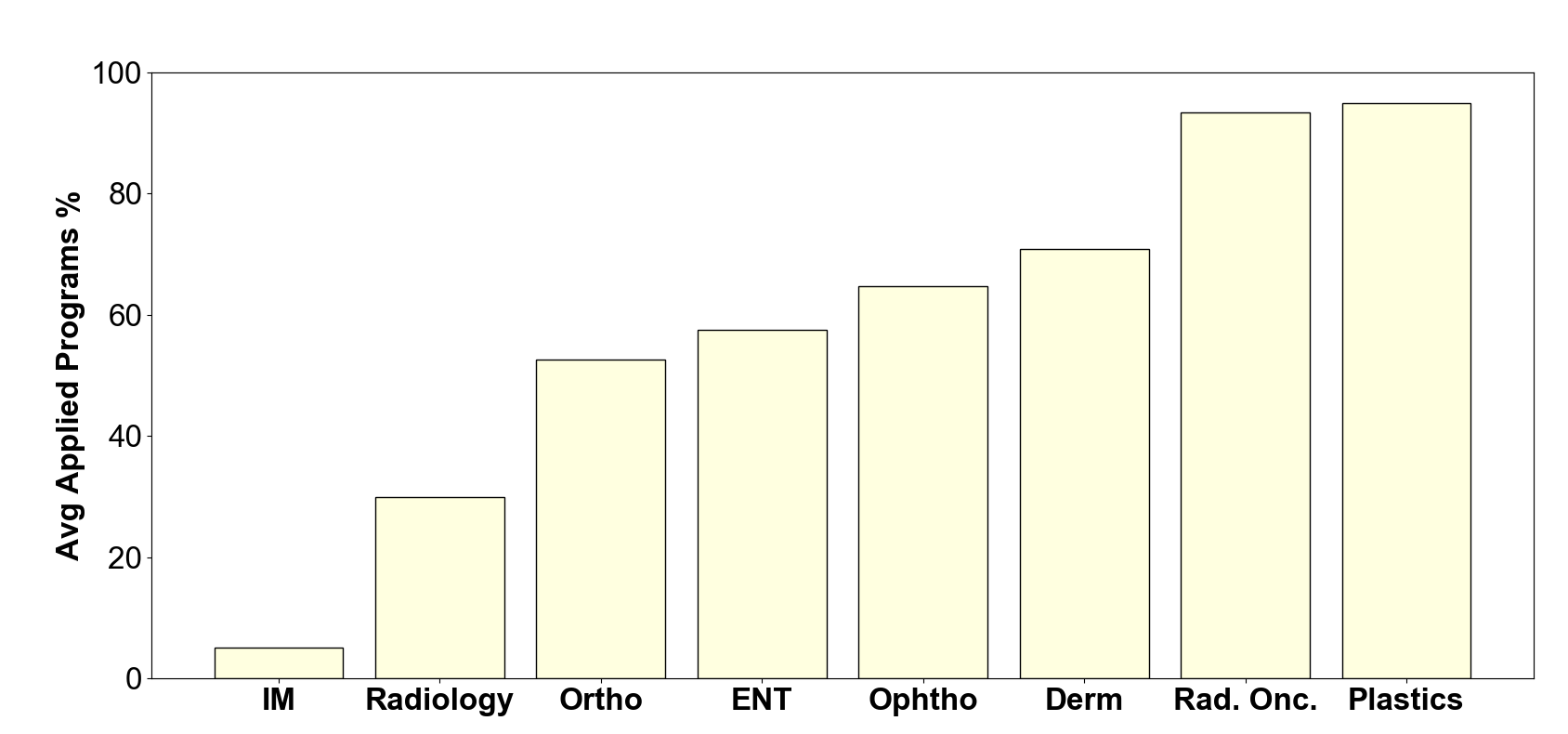} 
	\caption{Comparison of the number of applications as a percentage of all programs in 2019 for ophthalmology and National Resident Matching Program specialties internal medicine, radiology, orthopedic surgery, otolaryngology (ear, nose, and throat), dermatology, radiation oncology, and plastic surgery. }
	\label{fig:avgPerc_programs_applied}
\end{figure}

\section{Conclusion} \label{sec:summary}
The current Gale-Shapley algorithm for resident matching induces excessive applications and interviews for both applicants and programs because of its Bayesian Nash Equilibrium.
This Bayesian Nash Equilibrium increases costs for all participants while not changing overall payoffs.
\citet{wu2021inefficiencies} showed that rank lists can be significantly truncated for both applicants and programs without affecting the number of positions matched.

Another limitation of Gale-Shapley is the ordinal ranking of counterparties cannot account for relative preferences.
For example, an applicant can be expected to feel strongly about matching their top program versus matching their 5th choice program, but will they feel as strongly about matching their 11th ranked vs 15th ranked program?
Conversely, programs typically rank 10 applicants per spot \cite{wu2021inefficiencies}, but will a program with 6 spots put the same care and thought into ranking their 1st to 5th choice applicants versus the applicants ranked 55-60?

The lack of relative preferences and the increasing costs due to the Gale-Shapley's Bayesian Nash Equilibrium suggest the need for an overhaul of the current resident matching algorithm. 
A potential new algorithm could have applicants and programs relatively weigh potential partners up to a certain weight budget. 
Participants would still be able to apply and interview as many potential partners, but the weight budget would make additional applications and interviews less desirable as participants would dilute the weights they could give counterparties. 
Therefore relative weights would allow relative preferences,
while the weight budget would avoid Gale-Shapley Bayesian Nash Equilibrium.

\bibliographystyle{elsarticle-harv}
\bibliography{resident_matching_game_theory}






% \end{thebibliography}

\end{document}